POSITION PAPER

# Atomistic Modeling for Electrochemical Reactions

## From the State of the Art to a Strategic Roadmap for Future Research within *EU-CONCERT*

K. Doblhoff-Dier[1,*], B. Ballotta[2], L. Bonati[3], V. J. Bukas[4], F. Calle-Vallejo[5,6], C. S. Cucinotta[7], K. S. Exner[8,9,10], M. García-Melchor[6,11], L. Giordano[12], K. Honkala[13], N. Hörmann[4], G. Kastlunger[14], M.M. Melander[13], Zehra B. Öcal[15], A. A. Reka[16], S. Steinmann[17], M. Vandichel[18], J.-X. Zhu[19], G. Cicero[20]

* Corresponding Author: k.doblhoff-dier@lic.leidenuniv.nl

**Computational modeling plays a central role in advancing our understanding of electrochemical reactions and may thus guide the rational development of sustainable energy technologies. Despite significant methodological progress, achieving predictive and quantitatively reliable simulations of electrified solid–liquid interfaces under operando conditions remains a major challenge. Recently, the European COST Action *EU-CONCERT* (EUropean COllaborative Network on electroCatalysis for Efficient Renewable Technologies) was launched with the aim of advancing atomistic modeling of electrochemical reactions through systematic benchmarking and cross-validation. This position paper outlines the consortium's view of the current state of the art, identifies the priority methodological challenges to be addressed, and defines the benchmark tests that are urgently needed. This paper should serve as a guide to the community and a roadmap for the EU-CONCERT COST Action.**

Author affiliations: [1]Leiden Institute of Chemistry, Leiden University, PO Box 9502, 2300 RA Leiden, The Netherlands, [2]School of Physics, Trinity College Dublin, Dublin, Ireland, [3]Atomistic Simulations, Italian Institute of Technology, 16145 Genova, Italy, [4]Fritz-Haber-Institut der Max-Planck-Gesellschaft, Faradayweg 4-6, D-14195 Berlin, Germany, [5]Nano-Bio Spectroscopy Group and European Theoretical Spectroscopy Facility (ETSF), Department of Advanced Materials and Polymers: Physics, Chemistry and Technology, University of the Basque Country EHU, Avenida Tolosa 72, 20018 San Sebastián, Spain, [6]IKERBASQUE, Basque Foundation for Science, Plaza de Euskadi 5, 48009 Bilbao, Spain, [7]Department of Chemistry and Thomas Young Centre, Imperial College London, Molecular Sciences Research Hub, London W12 0BZ, U.K., [8]University Duisburg Essen, Faculty of Chemistry, Theoretical Catalysis and Electrochemistry, Universitätsstraße 5, 45141 Essen, Germany, [9]Center for Nanointegration (CENIDE) Duisburg-Essen, 47057 Duisburg, Germany, [10]Cluster of Excellence RESOLV, 44801 Bochum, Germany, [11]Center for Cooperative Research on Alternative Energy (CIC energiGUNE), Basque Research and Technology Alliance (BRTA), 01510 Vitoria-Gasteiz, Spain., [12]Department of Materials Science, University of Milano-Bicocca, 20125 Milano, Italy, [13]Department of Chemistry, Nanoscience Center, University of Jyväskylä, Finland [14]Department of Physics, Technical University of Denmark, Fysikvej 307, 2800 Lyngby, Denmark, [15]Sinop University, Faculty of Engineering and Architecture, Department of Environmental Engineering, Sinop, Türkiye, [16]Faculty of Natural Sciences and Mathematics, University of Tetova, Republic of North Macedonia, [17]CNRS, ENS de Lyon, LCH, UMR 5182, 69342, Lyon cedex 07, France, [18]Materials and Catalysis Modelling Group, Bernal Institute, University of Limerick, [19]Theory and computer-aided modelling of materials in energy engineering (IET-3), Forschungszentrum Jülich GmbH, 52425 Jülich, Germany, [20]Department of Applied Science and Technology, Politecnico di Torino, C.so Duca degli Abruzzi 24, 10129 Torino, Italy.

Electrocatalysis is central to a number of key electrochemical technologies for sustainable energy conversion and storage, including water electrolysis, $CO_2$ reduction, and fuel cells. The efficiency, selectivity, and stability of these processes are governed by complex phenomena at electrified solid–liquid interfaces, where atomic-scale understanding remains incomplete.

Computational modeling plays a crucial role in advancing and widening this understanding and enabling knowledge-based design strategies. Although we can currently look back at more than two decades of significant methodological advances, we still lack quantitative prediction capabilities for electrochemical reactions in operando. This sobering assessment emerges not only from the complexity of electrochemical systems, but also from a lack of understanding of the shortcomings and limitations of our models and methods.

To advance the field of computational electrocatalysis, benchmarking current and future approaches is paramount. In response to these challenges, the European COST Action *EU-CONCERT* (EUropean COllaborative Network on electroCatalysis for Efficient Renewable Technologies) was recently launched, bringing together about 250 researchers across Europe. The Action aims to advance atomistic modeling of electrochemical reactions by means of structured benchmarking, cross-method validation, and coordinated collaboration between theoretical and experimental groups.

Through close collaboration between computational and experimental research groups, the consortium will identify and study well-defined and as-simple-as-possible model systems. Additionally, collaborations between different theoretical and computational groups will target cross-method validation of different models and simulation approaches. Such coordinated interaction between groups (both theory-theory and theory-experiment) is a particular stronghold of the consortium and can be expected to yield new insights into the strengths and limitations of current methodologies as well as to stimulate the development of new theoretical and computational approaches.

The principal aims of the consortium are thus to provide insight into the limitations of computational methods used in the field and to address them. Through this bottom-up approach, we aim to provide well-founded and verified computational methods that will allow for reliable simulations of complex systems in the future. Addressing complex, real-world problems in electrocatalysis will follow once these foundational methodological challenges have been resolved.

In this position paper, we first describe our standpoints regarding various atomistic simulation approaches relevant to a computational description of electrochemical reactions, then we discuss challenges inherent to benchmarking against experiments. Finally, we summarize the state of the art, known limitations and challenges of various approaches in the appendix.

Throughout this position paper, we consolidate the viewpoints expressed during the first EU-CONCERT online meeting while articulating a clear vision for the consortium.

## Challenges in electrochemistry

Overall, we identify 4 main challenges in computational modeling of electrochemical reactions:

1) How do we effectively and accurately predict and simulate electrochemical interfaces under operando conditions?
2) How can we correctly account for the effect of applied potential and the resulting electric field in our simulations?
3) How does the (dynamic) electrolyte affect electrochemical reactions?
4) How can we (de)couple processes occurring on different time- and length-scales?

Within *EU-CONCERT*, it is our aim to eliminate any influence from point 1 as much as possible by focusing on atomically well-defined systems in both experiments and simulations. This way, the grand challenge of determining the operando interface structure can be postponed to a point where we know that we can reliably describe processes at charged interfaces under potential control and we focus on challenges 2 to 4 in the following and only briefly summarize methods to tackle challenge 1 in the state of the art description in the appendix.

# Atomic-scale approaches in electrochemistry

To address the atomic scale computational challenges in electrochemistry, various computational approaches are available. In the following, we sketch the consortium's current position on three different approaches that we consider particularly relevant in the context of our aims, namely:

1) Static DFT modeling under potential control based on implicit solvent models.
2) Simulating thermodynamic quantities and dynamics.
3) Machine learning interatomic potentials in electrochemistry.

## 1) Static DFT modeling under potential control and implicit solvent models

A major breakthrough in atomistic simulations of electrochemical processes was triggered by the development of the computational hydrogen electrode (CHE) referencing technique.[1,2] This minimalistic approach includes the effect of the applied potential a posteriori and can thus account for proton-coupled electron transfer (PCET) thermodynamics. At the same time, it ignores local effects due to the field/capacitive charging and partial electron transfer at the electrochemical interface that can play a role under realistic conditions.[3] The CHE method has proven highly informative in many cases,[4] is still considered a good first approximation, and remains key for electrocatalytic materials screening. However, for more detailed mechanistic studies, electronically grand canonical DFT[5–9] methods based on implicit solvation models are typically used. These methods are not restricted to describing PCET and allow for the inclusion of electric field effects.[10,11] Nevertheless, these methods have their own shortcomings. For example, they (i) struggle to provide a correct reference energy for protons in bulk solution, which is why they have sometimes been combined with the CHE approach[12,13] and (ii) face challenges[14] due to the approximative description of solvation,[15] leading to open questions regarding their reliability. This explains why, during the *EU-CONCERT* online meeting held on January 7, 2026, there was a clear call to better understand the reliability and applicability of the different constant potential models and methods available.

Implicit solvent models, as used in constant-potential DFT calculations, achieve a thermodynamic equilibrium description of the liquid by construction. A source of concern is, however, the degree of reliability in this description that can be or is currently achieved. Ideally, implicit solvent models should correctly predict the potential of zero charge (PZC), the double layer capacitance profile in presence and absence of different adsorbates/coverages, and solvation free energies. In practice, many common implicit solvation models struggle to achieve an accurate description of double layer quantities[16] and solvation free energies[17–19] and more involved models are often needed.[20] In fact, the **absence of transferable solvent parameters for dielectric continuum models** was identified as a main challenge for constant-potential DFT by the consortium.

To parameterize and benchmark solvation models for electrochemical interfaces, knowledge of the double layer capacitance, the PZC, and adsorbate solvation free energies are key. Unfortunately, these are difficult to access experimentally. For instance, the double layer capacitance (available e.g., for Ag(100)[21]) is often convoluted with adsorption events (see e.g., ref. [22]), and solvation free energies generally require an indirect approach[23] that has only rarely been taken and does not seem to allow separating electric field effects from solvent effects for polar and charged species. Reliable experimental measurements of the double layer capacitance (with and without adsorbates) and solvation free energies (of neutral and charged molecules) for a large set of electrode materials and structures are important to benchmark and develop continuum solvation models.

In addition to the parameterization of the solvation models, the level of theory used plays a role. Inclusion of beyond-mean-field effects (as in classical DFT,[24,25] or density-potential functional theory[20]) may allow for a better description of the capacitance and solvation energies, but the performance of such models for surface solvation or adsorption free energies has not been systematically addressed and, to our knowledge, no such methods are implemented in constant-potential methods.

A straightforward approach to go beyond a mean-field solvation is to use micro-solvation (the inclusion of a few water molecules and/or ions). Micro-solvation can be used as a stand-alone method (if used in conjunction with the CHE[26]) or in combination with implicit solvation models and constant-potential DFT. Compared to implicit solvation alone, micro-solvation has the advantage to capture localized and directional interactions such as hydrogen bonding, which can be critical for an accurate description of the reaction energetics. Micro-solvation models are highly portable, as they can easily adapt to low coordinated surface sites or nanoparticles,[27] and the inclusion of a few water molecules can lead to results similar to those obtained for water bilayers on flat surfaces.[28] Despite these advantages of micro-solvation, a rigorous yet rapid way to assess the number of water molecules to be included in the model is lacking and consecutive addition is needed.[26] Additionally, the lack of thermodynamic sampling, leading, e.g., to large energetic fluctuations when a hydrogen bond is broken is a source of concern.

To validate existing and new (implicit and explicit) solvation models, cross-validation remains a clear necessity. An example that was mentioned by the EU-CONCERT participants is that nonlinear dielectric continuum models are now readily available,[8,29] but due to a lack of benchmarks showcasing their potential advantages, the community has been slow in adopting them. Furthermore, clear guidelines are fully missing as to when dielectric continuum models fail and solvent descriptions including correlation effects such as those included in RISM,[30] or QM/MM,[31–33] become necessary. **This knowledge gap arises largely from the absence of systematic cross-method comparisons** designed to minimize confounding factors, such as different initial states, different reaction conditions, different DFT setups and exchange-correlation functionals, etc. **The design of cross-method comparisons is** therefore also **a primary target for *EU-CONCERT*.** Unfortunately, cross-method comparisons of constant-potential simulations are often impeded by solvation models only being available in specific codes. The Quantum Espresso solvation package ENVIRON[34] for example, encodes other methods than VASPsol,[35] VASPsol++,[29] or JDFTx[8] and the situation becomes even more complex when considering advanced solvation models such as RISM,[30,36,37] classical DFT,[25] models based on statistical field theory,[38–40] models including additional interaction parameters,[20,41] and QM/MM approaches. The EU-CONCERT consortium **encourages the development of an open-source, transferable library of solvation models** that can be interfaced with various codes, in analogy to libxc for exchange-correlation functionals.[42] We furthermore encourage developers of quantum chemistry software to cater towards such a development by providing the necessary interfaces.

Overall, the various solvation and constant-potential schemes (explicit electrolyte description + molecular dynamics, implicit solvation models, micro-solvation approaches, and combinations thereof) are all valuable as they come at different computational cost and have different strengths. A clear understanding of their limitations and risks is, however, important (see also the appendix). At the same time, we should not forget the limited accuracy of the underlying DFT calculations: even with a suitable constant-potential and solvation scheme, energetics obtained by DFT can easily be off by $\mathcal{O}(100)$ meV. Given the electrochemist's focus on electrochemical issues, DFT inaccuracies are often taken lightly. To bring more attention to this issue, a section on DFT errors and their relevance to computational electrocatalysis can be found in the appendix.

## 2) Simulating thermodynamic quantities and dynamics

An inherent challenge in electrochemical modeling is the need to account for thermodynamic sampling. This includes sampling of the configurations of a reactant, intermediate or product at the interface, but also sampling of the electrolyte configurations responsible for the solvation and the formation of the double layer. While adsorbate configurations can, in principle, be assessed via minimum hopping algorithms, the electrolyte generally needs a thermodynamic description that can be achieved through molecular dynamics (MD) simulations,[43–48] though alternatives exist and have been used in computational electrochemistry.[49] Encouraging qualitatively correct results have been obtained with DFT-MD for key quantities, such as the PZC[44,50,22,51] and the inner layer capacitance.[52–54] However, quantitative differences to experimental results can remain, which can be attributed to approximations in the exchange–correlation functional, limitations in the models (including the neglect of nuclear quantum effects),[49,55–57] and the short timescales considered.

In view of the different timescales for solvent reorganization and ion transport, as well as the dimensions of a double layer and realistic interfaces, immense sampling efforts are needed. Therefore, it seems plausible that machine learning (ML), QM/MM, and enhanced sampling approaches[58,59] will play an increasingly decisive role in computational electrochemistry to reach time- and length-scales unachievable with DFT-MD. At present, and as discussed in the next section, however, ML approaches are still underdeveloped for electrochemical applications and would benefit from better validation. **To benchmark ML-MD approaches, long-timescale (but small length scale) DFT-MD simulations and post hoc DFT validation[60] seem essential.** DFT-MD should thus not be viewed as "superseded" by ML-MD at the present time but should be considered as an important validation of ML approaches. **Only after establishing reliable ML-MD approaches, will we be able to address outstanding challenges**, such as scrutinizing electric double layer models,[61] determining ion distributions and concentration profiles.

Building on thermodynamic sampling, and going beyond static DFT, electrochemical modeling also critically depends on the evaluation of free energy profiles. Free energy profiles for electrochemical reactions obtained from DFT-MD simulations (with and without potential control) have become increasingly available in the literature in the last few years.[62–66] However, the authors' subjective view is that many results for reaction barriers obtained from (electrochemical) free energy sampling appear scattered and occasionally difficult to rationalize. **It is, at present, not clear whether this scatter in electrochemical free-energy sampling for barriers comes from human error, system settings, or inherent limitations of approximations made.** [Note though that there are clear indications for statistical limitations in some cases (compare e.g., ref.[63] to the immense sampling effort required in ref.[67]; or ref.[65] with the error bars obtained for much longer simulation times in ref.[68])]

To obtain reliable results from free energy calculations, the models used in the simulation of the actual system dynamics also require attention. For instance, there is growing interest in modeling the charge and potential fluctuations[69] in constant potential DFT-MD through (thermo-)potentiostats[70] but the influence of fluctuations on e.g. reaction barriers, solvent and reorganization dynamics, transport properties and autocorrelation functions, and capacitance has rarely been addressed.[71] Additionally, the results depend on the limitations of current (thermo-)potentiostat schemes, which rely on (unknown) relaxation timescales and, for MD simulations, often (though not always[72,73]) on hybrid explicit/implicit approaches.[74] The risks and limitations of this approach are discussed in the appendix.

Overall, systematic, cross-lab benchmarking of electrochemical barriers and system dynamics as obtained via direct or enhanced sampling MD are needed. Such tests should include systematic variation of the potential and must prove compatible with important thermodynamic principles. The complexity of the system should only be gradually increased (e.g., by avoiding explicit interactions of ions with reactants) and the scientists involved should strive for agreement in terms of interpretation of the results. This **agreement is** currently **lacking. Community efforts and cross-lab validations,** as common in fields such as astrochemistry, **are therefore much needed** in the field of computational electrochemistry.

## 3) Machine-learned interatomic potentials in electrochemistry

The opportunities for machine learning in electrochemistry are vast and range from fast sampling of energetics for (adsorbed) reactants, intermediates and products, to the acceleration of MD simulations targeting solvation effects and the structure of the electrochemical double layer. Each of these potential fields of application faces its own challenges: while the prediction of adsorbate energies seems primarily limited by the quality and amount of the input data, the quality of ML-accelerated simulations of the double layer seems more fundamentally influenced by the architectural assumptions of current ML interatomic potentials (MLIPs) - in particular their locality, their treatment of long-range electrostatics and the absence of explicit electronic degrees of freedom - as well as by the transferability of training data from DFT-MD or similar approaches.

To predict reaction energetics, most contemporary ML approaches rely on DFT data of adsorbate energies. These may be computed in gas phase or micro-solvation and by using standard DFT or constant-potential DFT (which requires the applied potential as additional input). Whichever method is used, MLIPs will always inherit the underlying DFT errors present in the training set.[75]

Available datasets for training MLIPs include the open catalyst datasets 2020,[76] 2022,[77] and 2025[78] (the latter of which includes solid-liquid interfaces with explicit solvent environment), ElectroFace[79] (solid-liquid interfaces) and BEAST DB[80,81] (grand-canonical, constant-potential DFT data). Transferability of so-trained MLIPs to transition states and reaction intermediates remains, however, questionable and active learning coupled to enhanced sampling seems unavoidable.[82,83]

When addressing the double layer and its dynamics, a fundamental issue arises from the locality assumption underlying most machine-learning interatomic potentials, where the energy is decomposed into short-ranged atomic contributions. In systems of electrochemical interest – including polar liquids, ionic materials and charged fragments – long-range electrostatics and collective polarization effects play a central role, and purely local models may misrepresent properties such as charge distributions and dynamics.[84] Although message-passing schemes can be used to enlarge the effective range of the interactions (receptive field), the correct description of the Coulomb interactions at the interface requires explicit modeling of long-range interactions.[84] While dispersion interactions can often be incorporated through semi-empirical corrections[85] or sufficiently large cutoffs, the treatment of long-range electrostatics and nonlocal polarization remains substantially more challenging. An excellent recent review on the topic can be found in ref.[86]

Charge-aware neural network potentials, such as third-generation high-dimensional neural network potentials[87] or latent Ewald summation (LES),[88,89] as well as approaches based on the prediction of Wannier Centers (DPLR),[90] incorporate explicit long-range electrostatics and can alleviate issues arising from missing long-range effects. Electronegativity-based schemes such as fourth-generation neural networks,[87] similarly enable long-range charge redistribution. However, all these schemes struggle with the simultaneous description of the qualitatively different long-range responses of the two phases at an electrochemical interface: collective metallic screening on the solid side and long-range dielectric polarization and solvent reorganization in the electrolyte. Capturing both effects consistently within a single classical or machine-learning framework is challenging and remains an open problem. Recent approaches[91–93] that tackle this issue are promising, but unfortunately remain ill-benchmarked, and add an additional layer of complexity. In short, **electrocatalytic systems**, where electron-conducting solids are coupled to ion-conducting liquids and charge flows across the interface, **continue to push both DFT and MLIPs to their limits,** motivating hybrid (e.g. QM/ML) or multiscale strategies that retain at least an implicit description of the electronic degrees of freedom.

## Benchmarking against experiment

Ultimately, only a successful comparison to experimental data may validate a computational method. Such a comparison must go beyond the use of scaling relations and descriptor-based approaches: although highly successful for rationalizing catalytic trends, these approaches do not always translate into quantitative agreement with experimental performance.[94–96] Therefore, in the context of computational electrocatalysis, **the direct comparison between predicted and measured capacitance, free energies, reaction rates and equilibrium potentials for ad-/desorption are central.**

The advantage of benchmarking against reaction rates is that these are reasonably straightforward to measure compared to double layer capacitances (which are often convoluted with adsorption capacitances), the potential of zero free charge (which may not even be accessible for a surface with a particular adsorbate coverage and is a somewhat ill-defined quantity in systems in which ions can partially transfer their charge), ion distributions (the detection of which requires X-ray techniques), or (solvation) free energies or entropies (which are hard to assess experimentally). The disadvantage of benchmarks against reaction rates, rate constants, and adsorption equilibria is that (i) small changes in $\Delta G^{\ddagger}$ culminate in large changes in the reaction rate due to the underlying exponential relation and (ii) the simulation of reaction rates may require the consideration of the full reaction network and mass transport effects. To simplify matters, one may consider trends rather than absolute values, or exploit pH and kinetic isotope effects which can probe the rate-determining step along complex pathways.[97] Alternatively, one can try to exploit an automated exploration of reaction networks, as already established in fields such as combustion and astrochemistry. These approaches reduce bias in mechanism selection and enable the identification of competing pathways that may otherwise be overlooked. Recent developments extend these concepts to catalytic and surface systems, allowing for the construction of large reaction networks and associated microkinetic models.[98,99] Application to electrochemical interfaces remains challenging, however, owing to the large configurational space, the presence of solvent and electrolyte, and the dynamic nature of the surface, an exploitation of ML–based approaches seems unavoidable.[100–103]

In any case, a successful experiment–simulation comparison must consider the electrochemical operando conditions, including surface structure, defect concentration, but also mass transport, (local) pH, as well as impurities in the cell. Capturing these conditions is inherently complex. It is therefore a major effort for the experimentalists involved in EU-CONCERT to select **systems that are highly controllable or well defined** in terms of surface structure, defect concentration, mass transport, pH, and impurities in the electrolyte. It is the vision of EU-CONCERT that well-controllable or well-defined **systems should be analyzed via cross-lab validation and results be compared for reproducibility before aiming for an experiment–theory comparison**.

## Conclusion

Overall, in this position paper, we have not only highlighted the target points of the theory workgroup of the COST-Action EU-CONCERT, but have also charted methods in computational electrochemistry, and summarized the state of the art and critical limitations. We hope to inspire computational electrochemists to work on the highlighted challenges and, in particular, to seek close interaction with other computational, experimental, and theoretical groups, allowing for benchmarking and cross-validation. The COST-Action EU-CONCERT (https://eu-concert.eu) provides a key platform for doing so. In line with the inclusiveness objectives of COST, the Action also offers valuable opportunities for Young Researchers and Innovators (YRIs) and researchers from Inclusiveness Target Countries (ITCs) to establish international collaborations, gain visibility within the community, and actively contribute to shaping the future of computational electrochemistry. We warmly encourage prospective members to apply through the COST-Action application form (https://www.cost.eu/actions/CA24103/).

## Acknowledgements

This article is based upon work from COST Action EU-CONCERT, supported by COST (European Cooperation in Science and Technology). The authors acknowledge helpful discussions with many of the EU-CONCERT members. Several members, as listed in the Supporting Information, have endorsed the paper. Large language models have been used to improve the phrasing on the sentence level and on the level of sentence coherence. All test improvements suggested by large language models were critically evaluated.

## Appendix

### State of the art, known limitations and challenges in electrochemical modeling

The following “state of the art” reflects, to a certain extent, the opinion of the authors, informed through the participants of the first online meeting of the COST Action EU-CONCERT.

### 1) Static DFT modeling under potential control and (implicit) solvation models

We focus here on static DFT calculations that achieve potential control via implicit solvation models. For other electrification schemes, we refer to a recent review, ref.[69]

State of the art:

For interface calculations, typical system sizes consider two-dimensional periodic slabs consisting of below 40 surface atoms and 2-5 subsurface layers. The relevance of defect sites and defective surfaces is increasingly recognized and evaluating them is considered good practice. As exchange-correlation functional, mostly PBE,[108] RPBE[109] or BEEF-vdW[110] are used, where the first is the overall most applied exchange-correlation functional for solids, RPBE provides accurate O, CO and NO adsorption energies on selected transition metals, and BEEF-vdW was fitted, amongst other things, to some adsorption energies.

In the following, we discuss different techniques typically used in computational electrocatalysis, including the computational hydrogen electrode approach, constant potential DFT, and micro-solvation. Finally, we also discuss the limitations of DFT as typically used in computational electrocatalysis.

- The **Computational hydrogen electrode** method, though often considered “old school”, still has relevant applications in computational electrochemistry. In particular, it is

- Still considered state of the art for electrocatalyst design and selection via (high-throughput) materials screening
- A relevant ingredient when modeling reaction networks,[104,105] as constant-potential DFT methods are (currently) incapable of accurately describing protons in a bulk-like environment.

*Known limitations:*

- Only applicable to PCET thermodynamics or, more generally, processes describable through closed thermodynamic cycles analogous to Born–Haber constructions. i.e., no partial electron transfer is possible. Therefore, it is also
    - Unable to describe, non-Nernstian effects/non-integer potential response[106]
    - Not intended for the direct calculation of electrochemical reaction barriers.
- Limitations stemming from the exchange-correlation functional (see below for details).

- **Constant-potential DFT or values corrected to a certain potential via a post processing correction[107] for barrier and mechanistic studies.**

  Most constant-potential DFT approaches rely on implicit linear dielectric continuum solvation models, which can differ quite widely in flavor and parameterizations. Widely used implementations are available via

  - ENVIRON[34] (QuantumEspresso)
  - VASPsol[35,112] and VASPsol++[29] (where VASPsol++ is largely considered to supersede VASPsol due to its enhanced stability and the availability of nonlinear models and solvent-aware implementations)
  - JDFTx[8]
  - GPAW/SJM[7]

  *Known limitations:*

  - Possibly incorrect representation of the capacitance, leading to an incorrect relationship between potential and surface electric field, possibly impacting energetics. This originates (mostly) from the solvent model.
  - Likely incorrect representation of the potential profile, originating (mostly) from the solvent model.
  - Limitations stemming from the exchange-correlation functional (see below for details)
  - Incorrect surface solvation energy
  - Correctly describing the initial/final state for charged reactants/products is close to impossible as this would require bringing the charged species into the bulk of the electrolyte, which is (a) typically not done (often the particle is only brought to the reaction plane) and (b) requires an accurate description of solvation for charged species as otherwise the overall reaction energetics will be impacted.

  *Unquantified risks:*

  - Static DFT models are (in conjunction with transition-state searches) also often used to extract kinetic information. There is some debate on whether constant potential or constant charge ensembles provide a more realistic representation when describing dynamic events.[113] We argue that in many situations equilibrium conditions are sufficient and transition-state theory is valid.[114,115] Under equilibrium conditions, in the infinite system size limit, both approaches are thermodynamically equivalent.[116,117]

*Required improvements:*

- High-quality solvation models, ideally, with transferrable parameterizations specific to solid-liquid interfaces.

- **Micro-solvation**

  Micro-solvation includes local solvation effects in static calculations with a limited computational cost. Examples include solvation of oxygenates on metal nanoparticles,[27,118] carbon-based, and nitrogen-based intermediates on metallic surfaces.[119,120]

*Known limitations:*

- Dependence of the results on the number of water molecules included in the model.
- Due to the fluxional nature of water clusters, an increasing number of almost isoenergetic local minima may be found as more water molecules are included.[28] As a result, a single minimum might not be representative of the solvation structure.

*Unquantified risks:*

- Inaccurate solvation energetics may arise from insufficient sampling of water configurations, leading to an incorrect description of enthalpic and entropic contributions.

Alternatively to performing calculations directly at constant potential, post-processing schemes can also be used. Constant potential DFT[5–8] and post-processing corrections based on the capacitance and work function change[107] are often considered thermodynamically equivalent. In practice, however, differences arise a) due to the approximations needed in both frameworks, and b) also from the varying geometries obtained under constant-charge and constant-potential conditions.[111] Constant-potential DFT is often conceived as easier to use.

- **Exchange-correlation functionals**

  While sophisticated exchange-correlation functionals are available for specific applications, the situation remains difficult whenever metal surfaces and molecules need to be well described simultaneously[121,122] or when computational cost plays a crucial role.

  The three most widely used exchange-correlation functionals in catalysis are arguably PBE,[108] RPBE,[109] and BEEF-vdW.[110] They are massively used because of their cost-effectiveness and because PBE and RPBE, as general purpose functionals, can be expected to show somewhat acceptable accuracy over a wide range of systems. While researchers are often familiar with the limited accuracy of DFT, a comprehensive error assessment is not yet a standard procedure in electrochemistry and benchmarks for molecule–surface interactions are mostly limited to a couple of adsorption energies[110,123–126] and even fewer reaction barrier heights.[121,122,127–129] Most benchmarks are done for molecules and mostly assess average performance of functionals.[130] With respect to electrochemistry, to date, the most widely known error is that of $O_2$, which is routinely corrected in oxygen evolution and reduction modelling.[1]

*Known limitations:*

- Using GGA (such as PBE, RPBE and BEEF-vdW), meta-GGA and even hybrid functionals involves large errors for the thermochemistry of numerous gas-phase molecules and ions.[75,131] However, the errors appear to be systematic and can be corrected via descriptor-based analyses or using automated procedures.[75,132]
- Gas-phase errors have direct impact on calculated reaction energies, equilibrium potentials and free-energy

diagrams,[75,133] surface and bulk Pourbaix diagrams,[131,134] and volcano plots.[135]

*Unquantified risks:*

- The DFT errors in adsorption energies are a convolution of molecular and solid-state errors, which both may be influenced by the molecule adsorbing.[136,137] To quantify them, abundant experimental adsorption energies are needed, that go beyond the currently available databases.[110] Recent joint computational-experimental approaches have started to address this matter,[138,139] but considerably more work is necessary and protocols should definitely be established.
- While the field of electrochemistry turns more to the determination of barriers, the large DFT errors[121,122,129] that can be incurred – at least for barriers of dissociative adsorption reactions – are often ignored. Benchmarking reaction barriers is arduous,[140] and in electrochemistry it is often convoluted with other difficulties related to the applied potential and electric field profile as discussed above.

## 2) Simulating thermodynamic quantities and dynamics

State of the art:

- ***Ab initio* molecular dynamics**

  Typical system sizes and length scales consider slabs of up to ~(6 × 6) surface-atom supercells (for metals; equivalent system sizes are used for oxides) and a several tens to a few hundred water molecules. Considerably smaller system sizes surface supercells (e.g., 3 × 3) are often still used though, in particular for thermodynamic integration of reaction processes. Simulation times are typically in the order of 10 to ~100 ps.[43,141,142]

  Charging is typically achieved by adding ions of one type[52,143] or, more generally, via an imbalance[144] between positive and negative ions in the electrolyte. Alternatively, DFT-MD simulations can also be coupled to mean-field solvation models to achieve potential control.[30,74,145] Such schemes will, however, bear risks/limitations as they (i) always describe the mean-field part of the solution in equilibrium, (ii) do not allow for (further) ions to penetrate the explicit electrolyte region and therefore likely do not reproduce the correct potential profile, and (iii) can show computational intricacies (see below). Finally, several other charging/electrification schemes have been suggested and used in literature, which we will not detail here (e.g., refs.[36,146–150]). We refer interested readers to a recent review[69] that, with the exception of refs.,[146,147] reviews these and other methods.

*Known limitations:*

- Small system size can lead to artifacts in the (liquid) structure of aqueous electrolytes and will limit the extent of the double layer to length scales typical of the inner layer.
- Limited time and length scales restrict achievable sampling efficiency and convergence. In particular, ion-density and free-energy profiles can be expected to be difficult to converge, and the use of enhanced sampling methods is essential. Yet, even with enhanced sampling, depending on the exact system and the enhanced sampling setup, considerable error bars may remain – even for timescales achievable only with ML-MD (compare Ref.[68] and Ref.[151]).
- Unknown bulk ion concentration, as "bulk" conditions cannot be achieved (at least not for ionic strengths <1 M) in the limited system sizes considered.
- Experiments fix the electrolyte activity and replicating this would require DFT-MD in the grand canonical ensemble for the ions and solvent. (Electronically grand-canonical approaches do not account for this).

- Treatment of charge/potential fluctuations and their influence on system dynamics. Systematic studies on the influence of these charge/potential fluctuations with respect to simulation cell size are needed.[71]

- Including nuclear quantum effects is challenging.[152]

- Unquantified potential effect of charge delocalization resulting from the use of GGA functionals.

*For coupled DFT-MD – implicit solvation:*

- Risk of penetration of the mean-field solvent into the explicit solvent region; difficulty of generating a "true" interface.

- At the implicit/explicit boundary, an (unphysical) potential drop can occur, limiting the meaningfulness of the value of the applied potential.

- Charged species cannot rigorously be brought into the "bulk" during free energy integrations (at least not when kept in the explicit region).

- For limited system sizes, neither constant potential, nor constant charge implementations will capture the correct charge and potential fluctuations that should occur in the electrolyte region close to the interface. Thermo-potentiostats[70] alleviate this to a certain extent but require the choice of a relaxation time. Additionally, today's approaches are limited to a single (dominant) relaxation process/time constant. The influence of these fluctuations is largely uncharted today.

- The implicit solvent will always remain in equilibrium, even if a (fast) reaction occurs (in the explicit region). While this can be simulated by describing the full reaction dynamics in a fully *ab initio* description, simulations that are coupled to a potentiostat will not correctly describe the state of the electrolyte. Thermo-potentiostats[70] can drive the system back into equilibrium, but (i) will suffer from the same limitations as mentioned above, and (ii) will not necessarily correctly capture the dynamic response of the system. We note that, in many cases, a non-equilibrium description will not be required.[114,115]

## 3) Machine-learned (ML) interatomic potentials for electrochemistry

State of the art:

**a) Use of off-the-shelf ML interaction potentials (MLIPs)**

A broad range of MLIPs seems to be used by the community, including ACE, MACE, DEEP-MD, NEQUIP and many more. In the current state of the art, off-the-shelf MLIPs are mostly used to extend timescales (or sampling) while keeping system sizes that are similar to those used in DFT-MD simulations. This is not an inherent limitation though (see e.g., refs.[92,153]). When increasing the unit cell size, care should be taken, though, that relevant long-range effects are still adequately accounted for in the production runs (see below). In this regard, we expect models that make use of explicit Coulomb interactions or nonlocal descriptors (see sections (d) and (e) below) to show comparative advantages.

Off-the-shelf MLIP frameworks, which typically do not account for explicit Coulomb interactions, charges or applied potential, can still be trained at a particular applied potential (or electric field). This can be achieved by training on data obtained via a mixed explicit/implicit solvation scheme or by charging via a homogeneous background.[153] Alternatively, the MLIP can be trained on fully *ab initio* data with an ion imbalance in the electrolyte, which also allows one to control the surface charge. In the latter case, training can even be performed for a mixture of surface charges. However, as shown and discussed in ref.[84], training at a given surface charge seems preferrable for typical system setups used today. To efficiently train multiple MLIPs corresponding to different charge values, transfer learning may allow considerable learning speedup.[154]

*Known limitations:*

- Missing long-range interactions (e.g., Coulomb interactions; van der Waals interactions) beyond the cutoff radius/receptive field.
- Lack of databases for solvated and electrified interfaces.

*Unquantified risks:*

- Inaccuracies in the models obtained can be hard to detect. Root mean-square errors for a test set are often insufficient to detect possible issues. (This holds for all MLIP-types discussed in the following)
- Possibly wrong/compromised learning due to the lack of Coulomb interactions as important physical effects.
- The missing/incorrect description of the effect of electric fields caused by species outside the cutoff radius/receptive radius.

In addition to these limitations and risks that pertain to the MLIPs themselves, many limitations and risks named above for DFT and DFT-MD calculations in electrochemistry transfer to the MLIPs through the training schemes. However, due to the significant speed-up and the (hopefully relatively limited) training data, other problems (e.g., limited timescales, limited system size, and use of GGA functionals in DFT) can potentially be avoided when using MLIPs.

**b) MLIPs using the potential or surface charge as additional input during learning and predicting**

When trained via implicit electrolytes, the surface charge[49,68,155] (or potential[156,157]) can be used as an additional descriptor. This allows the MLIP to learn predictions for varying surface charges (potentials). (Electronically grand-canonical) energies can then either be predicted directly,[68,155] or via an expansion of the energy and the knowledge of the (potential dependent) local charges predicted from an MLIP.[156] If the surface charge is used as an additional descriptor for constant-potential descriptions, the work function or the chemical potential of electrons $\Phi = \frac{dE}{dQ}$, where $E$ is the (free) energy and $Q$ the free charge, must also be predicted via the MLIP.

*Known limitations:*

- Due to the rotational symmetry of the descriptors typically used, this approach is expected to break down if species reside outside the receptive field of the surface, as they should lose the “sense of direction”. This is related to the limitations and risks listed for the off-the-shelf MLIPs. One may expect these models to work best for interfaces without explicit solvent or with a limited amount of explicit solvent.
- Similarly, if charged species reside outside of the receptive field of the metal, charge predictions for the metal atoms can be expected to be unreliable. This again suggests that these models are most suited in circumstances where the amount of explicit solvation is rather limited.
- Additionally, the missing long-range interactions will limit predictability for systems in which long-range effects in the bulk are relevant.

- One can expect that the approach should break down for inhomogeneous systems where, e.g., two surfaces with different PZC are mixed.

**c) MLIPs trained at one potential but allowing predictions at other potentials through first or second order corrections**

While most MLIPs must be trained explicitly at the potential desired, the response analysis in $z$-orientation (RAZOR) model[158] makes use of a second-order expansion to predict energies outside the training range (i.e., away from an applied surface charge $Q_0$). This is achieved through a charge + local-descriptor-based prediction of "local" contributions to the dipole polarization along the surface normal and the first derivatives of the atomic forces with respect to the applied surface charge. Similar learning and interpolation advantages likely hold true for ref.,[156] which uses a related approach, but starting from the electronically grand-canonical energy.

*Known limitations:*

- Same as for (b)

*Risks:*

- The concept of "local" contributions to the polarization may break down in the presence of ions. Further testing of the approach for such applications is needed.

**d) MLIPs including explicit Coulomb interactions**

MLIPs can be made charge-aware by predicting charges[68] of the constituents or the location of Wannier centroids[90] based on the local environment. These charges can then be used to compute Coulombic interactions explicitly.

At metal interfaces, however, the charge is not necessarily defined by the local environment. To overcome this problem, ec-MLP[91] was developed, which combines the prediction of Wannier centroids from the local environment (effectively capturing charge and short-range polarization) with a Siepmann-Sprik-type[159] response of the metal electrode. The same approach has recently been used to couple metallic response to latent Ewald summation (LES)[93] and a charge-aware approach based on Bader-basin-centroids.[92] The advantage of these approaches is that they allow fluctuating charges on each atom and thereby should, e.g., enable the discharge of an ion upon adsorption.

*Known limitations:*

- Although constant-potential simulations are possible through the two-electrode setup, one should not forget that the simulations are still inherently canonical in the atomic/ionic species. Also, with two electrodes, it may be difficult to fix the (absolute) potential of the "working" electrode against a reference electrode and not just the potential between reference and counter electrodes.
- Due to the specific Wannier centroid approach used, ec-MLP[91] is restricted to constant (total) per-atom charges. This is not a fundamental issue, however.

*Unquantified risks:*

- Unknown reliability of the Siepmann-Sprik-type description of the metallic response.
- Sensitivity of the results to the underlying model parameters in the Siepmann-Sprik model.
- The description of nuclear and electronic charges as localized may prove insufficient.
- The transfer of charge from the electrolyte (well localizable via Wannier approaches) to the metal may prove challenging for these methods.

**e) MLIPs that use attention to account for long-range interactions**

To introduce long-range interactions, so-called attention schemes have also been proposed. This is implemented, for example in SpookyNet.[160] In principle, this concept can be combined with an explicit description of Coulomb interactions through the prediction of (long-range aware) atomic charges.[68]

## 4) Reaction networks, mass transport, and reaction kinetics

State of the art:

- Reaction networks constructed based on chemical insight
- Inclusion of potential and pH effects based on constant-potential DFT
- Mainly in fields other than electrochemistry: Automated exploration of reaction mechanisms.[98–100,100–102]
- If mass transport plays a role, it should be coupled to the reaction kinetics, as often done in literature.[161] Mass transport can be captured via analytical approximations, as done in ref.,[161] or through numerical simulations based on the Poisson-Nernst-Planck equations.[162]

*Known limitations:*

- Oversimplified surfaces (e.g., single-crystalline); limited configurational sampling
- Solvent effects often treated approximately
- Potential and pH effects often treated approximately; e.g., Butler-Volmer-type kinetics with constant transfer coefficients are often employed and local electrochemical activity of species is obtained through simplified mass-transport equations.

*Risks:*

- An accurate treatment of mass transport must include buffering effects and can depend sensitively on the exact parameterization.[163]
- Hidden pathway bias when relying on hand-built reaction networks
- Misidentification of the operando structure; "overlooking" active sites
- Overlooking mass-transport effects.

## 5) Assessing operando structures

Real electrochemical surfaces and interfaces rarely resemble ideal single-crystal surfaces: they undergo reconstruction, dissolution, oxidation, reduction, facet evolution, defect formation, and dynamic restructuring under applied potential – a process that can even depend on the operando electrolyte environment (including local electrochemical activity of protons).[164–166] Often the relevant catalyst surface is "guessed" based on experimental insight. Modern computational approaches aim to determine the relevant active catalyst surface via (surface) Pourbaix diagrams[167,168] and consider complexities such as the active site distribution. However, a catalyst is generally not at thermodynamic equilibrium, and cycling history and mass transport can play a role. In the past, out-of-equilibrium DFT-MD simulations have been used to rationalize dissolution processes[169] and very recently, large-scale MD simulations based on MLIPs have been exploited to unravel electrode restructuring.[170,171] While these simulations have proven insightful, they only cover a small part of the picture. Overall, we see a need for data-driven approaches towards identifying active site distributions and catalyst reconstruction.[172,173] Besides, clearer efforts towards understanding the effect of different active sites on catalysis are needed.[174]

State of the art:

- Consideration of operando catalyst structure based on insight from experiment.
- Computational Pourbaix diagrams:[175,176] Surface and bulk Pourbaix analyses provide thermodynamic stability maps under

electrochemical conditions and help identify relevant surface terminations,[177] including oxides, hydroxides,[178] and hydrides,[171,179] together with subsurface structures.[180] These approaches have been applied to various catalytic substrates including metals, oxides, 2D materials, and nanostructures.[181]

*Known limitations:*

- Neglect dynamic phenomena and kinetic barriers
- When considering (surface) Pourbaix diagrams, mass transport limitations affecting the local electrochemical activity of protons are generally neglected,[168] potentially leading to an incorrect prediction of the surface state.[164]
- Surface Pourbaix diagrams often neglect surface heterogeneity and are not generally devised to predict surface reconstruction, which is important to enhance catalyst durability[182,183]
- Results often depend crucially on the exchange-correlation functional leading to some phases not being visible or of appreciably different sizes in the potential vs pH plane.[131,134]
- Joint experimental and computational studies have shown that surface Pourbaix diagrams might only be a fair starting point to elucidate the actual, more complex structure and degradation pathways of electrocatalysts under reaction conditions.[182,183] A feedback loop between theory and experiments is highly recommended in this area.

# Supporting Information: Atomistic Modeling for Elec-tro-chemical Reactions

K. Doblhoff-Dier, B. Ballotta, L. Bonati, V. J. Bukas, F. Calle-Vallejo, C. S. Cucinotta, K. S. Exner, M. García-Melchor, L. Giordano, K. Honkala, N. Hörmann, G. Kastlunger, M.M. Melander, Zehra B. Öcal, A. A. Reka, S. Steinmann, M. Vandichel, J.-X. Zhu, G. Cicero

Throughout the position paper, we consolidated the viewpoints expressed during the first online meeting of the COST Action EU-CONCERT while articulating a clear vision for the consortium. EU-CONCERT members had the possibility to endorse the paper thereby showing their involvement in the original discussions as well as their support of the viewpoint. We provide a list of endorsers from the COST Action EU-CONCERT below. The information printed was provided by the endorsers in an endorsement form.

Çağla Akkol; Karadeniz Technical University, Department of Chemistry

Alexander Bagger; Technical University of Denmark

Oğuz Bayındır; Istanbul Health and Technology University, Türkiye

Otavio Beruski; Hyet Hydrogen, The Netherlands

Michele G. Bianchi; Politecnico di Torino

Conor Brennan-Pollak; Trinity College Dublin, Ireland

Hasan Celik; University of Cukurova, Turkiye

Hilal Demir Kivrak; Eskisehir Osmangazi University, Eskisehir Tukiye

Francesca Fasulo; University of Naples Federico II

Didac A Fenoll; CIC energiGUNE, Spain

Arianit Withdrawal Gashi; University for Business and Technology -UBT

Ralph Gebauer; ICTP, Italy

José R. B. Gomes; CICECO-Aveiro Institute of Materials, University of Aveiro, Portugal

Nitish Govindarajan; NTU Singapore

Axel Groß; Ulm University, Germany

Hajar Hajar Hajar Mahmoudi; Istituto di Scienze e Tecnologie per l'Energia e la Mobilità Sostenibili (STEMS)

Ehsan Hajialilou; IMDEA Energy, Madrid, Spain

Joakim Halldin Stenlid; Department of Chemistry and Chemical Engineering, Chalmers University of Technology

Hendrik H. Heenen; Fritz-Haber-Institut der Max-Planck-Gesellschaft

Matej Huš; National Institute of Chemistry

Vladislav Ivanistsev; University of Latvia

Faruk Karadag; Türkiye

Dogan Kaya; Department of Physics, Faculty of Science and Arts, Department of Advanced Materials and Nanotechnology, Institute of Natural and Applied Sciences Cukurova University, Adana, Turkey

Mehmet Fatih Kaya; Erciyes University, Energy Systems Engineering Dept., Heat Eng. Division, Kayseri, Türkiye

Ziad Khalifa; The British University in Egypt, Egypt

Karolina Kordek-Khalil; Wrocław University of Science and Technology

Raju Lipin; University of Limerick, Ireland

Mirtha A. O. Lourenço; CICECO – Aveiro Institute of Materials, University of Aveiro, Portugal

Milan Z Momčilović; University of Niš, Serbia

Mohammed Azeezulla Nazrulla; National Institute of Chemistry

Aligholi Niaei; Türkiye

Igor A Pašti; University of Belgrade - Faculty of Physical Chemistry, Serbia

Anup Paul; Portalegre Polytechnic University, Portalegre, Portugal

Daniela Polino; SUPSI, Lugano

Alvaro Posada-Borbón; Chalmers University of Technology, Sweden

Michele Re Fiorentin; Politecnico di Torino, Torino, Italy

Karsten Reuter; Fritz-Haber-Institut der MPG

Francesca Risplendi; Department of Applied Science and Technology, Politecnico di Torino

Kevin Rossi; TU Delft

Adriano Sacco; Istituto Italiano di Tecnologia

Esther Santos; APRIA SYSTEMS SL

Nicola Seriani; ICTP, Italy

Setareh Orangpour; Germany

Anton Stasyuk; INMA (CSIC-UNIZAR)

Muhammad Umer; University of Limerick

Francesc Viñes; University of Barcelona, Spain

Alexander D. von Rueden; Fritz Haber Institute of the Max Planck Society, Germany

Robin J. White; Luxembourg Institute of Science & Technology

Cem Burak Yildiz; Bartin University, Türkiye